\documentclass[fleqn,10pt]{wlscirep}
\usepackage[utf8]{inputenc}
\usepackage[T1]{fontenc}
\title{Oculomotor trajectory mapping on body as an effective intervention to enhance attention}

\author[1,*]{Songlin Xu}
\author[1]{Xinyu Zhang}
\affil[1]{University of California San Diego}

\affil[*]{soxu@ucsd.edu}

\begin{abstract}
Increasing individuals' awareness of their body signals can lead to improved interoception, enabling the brain to estimate current body states more accurately and promptly. However, certain body signals, such as eye movements, often go unnoticed by individuals themselves. This study aimed to test the hypothesis that providing eye-movement-correlated tactile feedback on the body enhances individuals' awareness of their attentive states, subsequently improving attention. Our results demonstrate the effectiveness of such feedback in redirecting and enhancing attention, particularly in the presence of distractions during long-duration tasks. Additionally, we observed that people's gaze behaviors changed in response to the tactile feedback, suggesting an increased self-awareness of current eye movements and attentive states. Ultimately, these changes in gaze behaviors contribute to the modulation of attentive states. Our findings highlight the potential of eye-movement-correlated bodily tactile feedback to increase individuals' self-awareness of their eye movements and attentive states. By providing real-time feedback through tactile stimuli, we can actively engage individuals in regulating their attention and enhancing their overall performance.
\end{abstract}
\begin{document}

\flushbottom
\maketitle
%
%
\thispagestyle{empty}

Attention is one of the most fundamental cognitive processes, yet sustaining focus over time poses a persistent challenge \cite{corbetta2002control}. 
To mitigate this challenge, prior research explored a variety of intervention technologies such as neurofeedback \cite{marzbani2016neurofeedback,bu2021novel,patil2022neurofeedback} and interoceptive illusion \cite{blanke2012multisensory}. These interventions are designed to enhance users' self-awareness of their behaviors or physiological signals, encompassing vital metrics like brain activity \cite{faller2019regulation} and heart rate \cite{iodice2019interoceptive}, thereby facilitating self-regulation. Nevertheless, the acquisition of such signals necessitates specialized apparatuses which are unwieldy for day-to-day routines. In addition, the signal variations may not provide immediate insights into users' cognitive states.

Alternatively, awareness of one's gaze patterns can serve as a biofeedback signal \cite{gregori2016assessing}, readily available through ubiquitous webcams in working or living environment. For instance, when individuals are tasked with maintaining focus on the center of a screen to complete a given objective, there is a propensity for attentional lapses, resulting in inadvertent visual exploration \cite{unsworth2016pupillary}. In such scenarios, providing individuals with real-time feedback pertaining to their gaze patterns can effectively apprise them of their unconscious ocular movements, consequently redirecting their attention towards the designated zones on the screen \cite{toreini2020using}.  
However, 
using the conventional auditory/visual biofeedback channels to deliver gaze signals, as proposed in prior works \cite{balgera2010synchronization}, may introduce additional distractions and cognitive load for individuals.

\begin{figure*}
\centering
\includegraphics[width=1\linewidth]{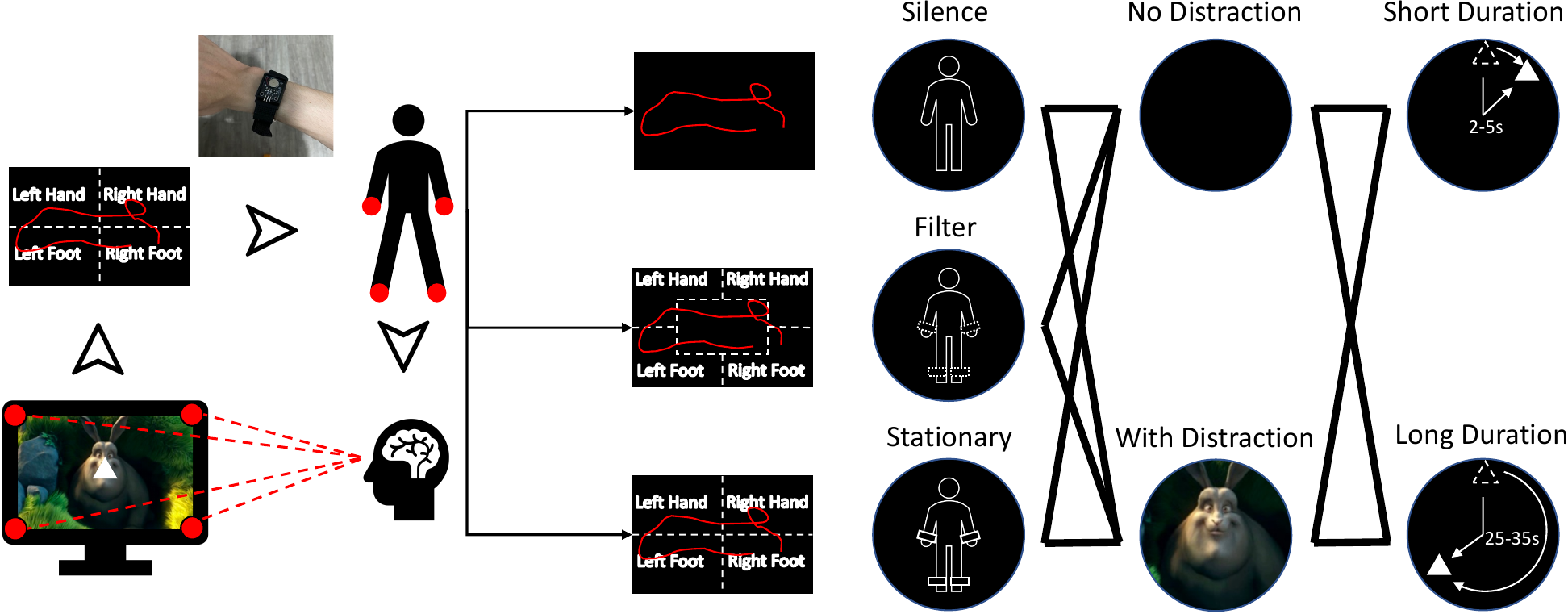}
\caption{Left: Eyerofeedback system. Users' eye movements are mapped to one of the four screen sub-areas, which will trigger the tactile feedback on the corresponding body location. Right: Study design illustration: 3 feedback type (\textit{silence}, \textit{stationary}, \textit{filter}) $\times$ 2 duration (short/long) $\times$ 2 condition (with/without distraction). Long duration condition is more difficult since users have to maintain their focus for a longer time.}
\label{f1}
\end{figure*}

In this study, we propose a novel approach that utilizes tactile feedback to establish a mapping between gaze patterns and the human body. 
As illustrated in Fig.~\ref{f1}, individuals receive vibration-based tactile stimuli on either their wrists or ankles, depending on the direction of their gaze. For instance, if an individual stares at the upper left of the screen, she will receive tactile feedback on the left wrist (more details in Materials and Methods).

Drawing inspiration from the concepts of neurofeedback \cite{marzbani2016neurofeedback} and interoceptive illusion \cite{blanke2012multisensory}, we put forth the hypothesis that \textit{this tactile bodily gaze mapping, referred to as \textbf{eyerofeedback}, has the potential to enhance user attention, even in the presence of distractions, by fostering heightened self-awareness of eye movements}.

To test our hypothesis, we recruited participants (N = 26) to finish an attention task (Three-Choice Vigilance Task: 3CVT) \cite{meghdadi2021eeg}, where participants were required to maintain their focus on the screen center and press corresponding keys on the keyboard when different shapes appeared. Each participant underwent 12 sessions (Fig.~\ref{f1}), which encompassed a combination of 3 feedback types (\textit{silence}, \textit{stationary}, \textit{filter}) $\times$ 2 duration settings (short/long) $\times$ 2 conditions (with/without distraction). 

This experimental design was based on our observations and existing literature \cite{mendoza2018effect}, which suggested that individuals may exhibit varying attentional performance in the presence or absence of external distractions and under different durations of attention. 
In the distraction condition, a movie \cite{WinNTBig} was presented in the background while participants completed the attention task, potentially inducing absent-mindedness earlier than in the no-distraction condition. The different durations refer to the varying time intervals at which the shapes appeared (short: 2-5 seconds, long: 25-35 seconds). A longer duration imposes a higher level of difficulty to the attention task \cite{posner1971components}. 
Regarding the feedback types, ``\textit{silence}'' served as the control setting, where no tactile feedback was delivered. The ``\textit{stationary}'' condition delivers tactile feedback corresponding to the participants' real-time eye movement directions, thus implementing the concept of \textbf{eyerofeedback}. The ``\textit{filter}'' condition represents a variant of eyerofeedback, in which tactile stimuli were only triggered when participants' eye movement distance exceeded a certain threshold (more details in Appendix).

Finally, our hypothesis posits that eyerofeedback exerts a substantial influence on user attention, particularly during long-duration sessions under distraction. In such sessions, individuals are more prone to experiencing lapses in attention, making eyerofeedback a crucial factor in regulating and redirecting their focus. We anticipate that the effects of eyerofeedback on attentional control will be notably pronounced under such circumstances.

\section*{Results}

We conducted a repeated-measures ANOVA to analyze the results of the aforementioned 12 sessions. To account for variability of participants' baseline performance, each metric mentioned below was normalized to have a zero mean and unit standard deviation within each participant.

\subsection*{Eyerofeedback Improves Human Attention}
\label{sec: R1}
\begin{figure*}
\centering
\includegraphics[width=1\linewidth]{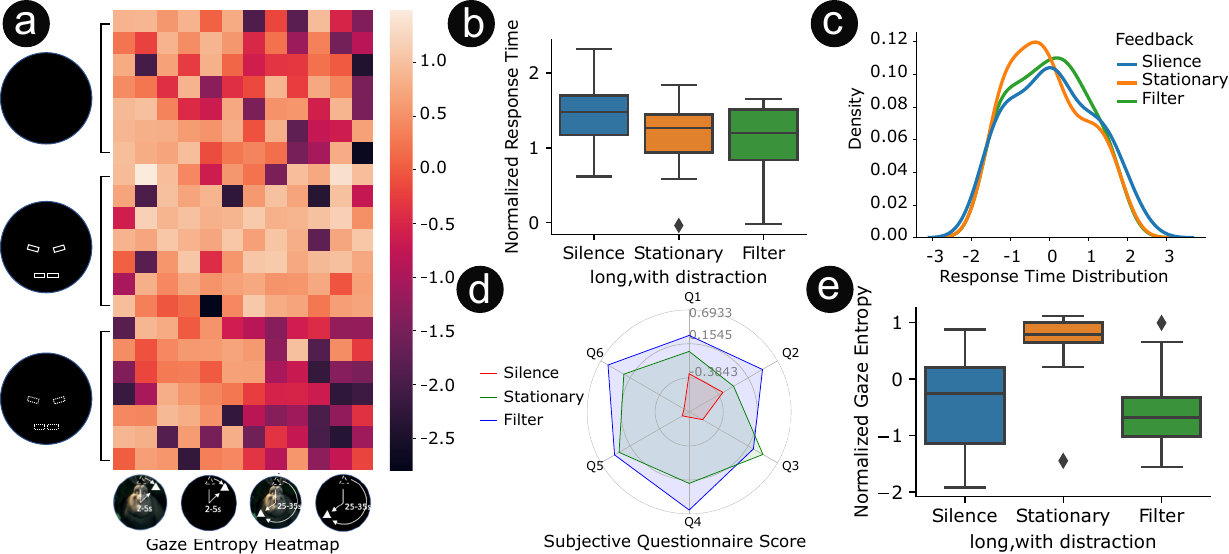}
\caption{(a). Gaze entropy heatmap for all participants. Each cell represents the gaze entropy of one participant in one of 12 sessions. (b). Box plot for the normalized response time across three feedback types of all participants over long duration with distraction condition. (c). Distribution plot for the response time across three feedback types of all participants over all duration and distraction conditions. (d) Radar plot to show the average score of each subjective question across three feedback types of all participants.
Questionnaires: To what extent you: can focus on the task (Q1), can focus on the screen center (Q2), feel distracted due to the feedback (Q3), think the feedback could improve your attention (Q4), feel your performance is affected by the feedback (Q5), hope the feedback could be used in your daily life (Q6).
(e). Box plot for the normalized gaze entropy across three feedback types of all participants over long duration with distraction condition. 
}
\label{f2}
\end{figure*}

In the attention task, participants were instructed to prioritize accuracy. Response time and the number of missed trials were used as metrics to quantify their performance. Our findings revealed that eyerofeedback significantly improved human attention compared to the control settings, particularly in more difficult (long duration) settings. 
Specifically, the ANOVA showed a significant main effect of feedback type on response time (F2,40 = 3.3135, P = 0.0466 $<$ 0.05) (Fig.~\ref{f2}(b,c) and Appendix Fig.~S2). 
There was also a significant interaction between feedback type and task duration (F2,40 = 3.8472, P = 0.0296 $<$ 0.05), indicating larger differences between feedback conditions in the long duration trials compared to short.
In addition, we observed a 
significant difference on response time across feedback types during long duration with distraction (F2,40 = 3.466, P = 0.0409 $<$ 0.05) and without distraction (F2,40 = 3.7076, P = 0.0333 $<$ 0.05). 
Further pairwise comparisons within the long duration setting demonstrated significant differences in response time for 
\textit{stationary} feedback compared to \textit{silence} without external distraction (\textit{stationary}: -0.5468 ± 0.3768 (mean ± SD) vs. \textit{silence}: -0.1726 ± 0.5523, P = 0.0367 $<$ 0.05), as well as for \textit{filter} feedback compared to \textit{silence} with external distraction (\textit{filter}: 1.1110 ± 0.4449 vs. \textit{silence}: 1.4790 ± 0.4096, P = 0.0243 $<$ 0.05). These results indicate that eyerofeedback has the potential to accelerate human response time and enhance attention levels.

The ANOVA also revealed a significant main effect of feedback type on the number of missed trials (F2,40 = 4.3364, P = 0.0197$<$0.05). However, follow-up tests showed the effect of feedback on missed trials varied across conditions. Specifically, feedback type only impacted missed trials in the long duration trials with distraction (F2,40 = 3.6115, P = 0.0362$<$0.05). No significant differences between feedback types emerged for missed trials in the short duration trials or long duration without distraction (P $>$ 0.05).
Furthermore, we did not find any significant differences in accuracy for feedback type across all four combinations of two duration $\times$ two distraction conditions (P $>$ 0.05). This result is reasonable considering that participants were instructed to prioritize accuracy over response time during the task.

The varied effects of eyerofeedback across conditions may be attributed to differences in baseline task performance. Response times were faster and fewer trials were missed in the less demanding short duration condition compared to the long duration condition. Additionally, the presence of external distractions led to slower response times and more missed trials than no-distraction conditions. Thus, eyerofeedback conferred larger benefits when baseline attention performance was lower due to greater attentional demands from long durations or external distractions.
This interpretation is supported by the significant differences in response time between the two duration conditions 
(F1,20 = 118.1033, P $<$ 0.001) and between the two distraction settings (F1,20 = 921.3792, P $<$ 0.001). Furthermore, significant differences were observed in the number of missed trials between the two duration conditions (F1,20 = 19.0266, P = 0.0003 $<$ 0.001) and between the two distraction settings (F1,20 = 9.967, P = 0.005 $<$ 0.01). These findings substantiate our hypothesis that eyerofeedback holds promise to improve individuals' attention performance by reducing response time and the occurance of missed trials. Nonetheless, these benefits varied as a function of task demands, emerging to a greater extent when sustained attention was challenged by longer durations or external distractions.

\subsection*{Subjective Feelings Reveal Eyerofeedback Effect on Human Attention}
\label{sec: R2}

Subjective experiences were assessed via a questionnaire administered after each session (Fig.~\ref{f2}(d) and Appendix Fig.~S3).
Participants reported heightened attention to the task (Q1) and screen center (Q2) with \textit{filter} versus \textit{silence} feedback. This was evident in both the short no-distraction (Q1: F2,40 = 6.3132, P = 0.0041; Q2: F2,40 = 6.9047, P = 0.0027) and long distraction conditions (Q1: F2,40 = 5.6305, P = 0.007; Q2: F2,40 = 8.0503, P = 0.0012). Across all conditions, \textit{stationary} and \textit{filter} feedback enhanced perceived attention (Q4) compared to \textit{silence} (P $<$ 0.001). However, eyerofeedback increased feelings of distraction (Q3) versus \textit{silence} (P $<$ 0.001). Participants also indicated performance was impacted by eyerofeedback (Q5) and expressed desire to use it daily (Q6) (both P $<$ 0.001). Together, these subjective reports offer compelling evidence that eyerofeedback heightens attention and gaze focus, thereby enhancing performance.

The participants' subjective ratings help explicate the differing attentional effects of \textit{stationary} versus \textit{filter} eyerofeedback across conditions. Without distraction, \textit{stationary} feedback heightened perceived attention compared to \textit{filter}. However, with distraction, \textit{filter} elicited higher attentional ratings than \textit{stationary}. This aligns with the objective performance data showing \textit{stationary} feedback expedited response times without distraction, while \textit{filter} feedback conferred benefits with distraction. The discrepancy suggests distractions may amplify \textit{stationary} feedback effects on anxiety, whereas \textit{filter}'s conditional tactile triggers may mitigate distraction-prompted anxiety. Supporting this interpretation, participants reported significantly greater belief in (Q4) and desire to use (Q6) \textit{filter} versus \textit{stationary} feedback (both P $<$ 0.05). By alleviating anxiety, \textit{filter} eyerofeedback may optimize attentional modulation amidst distractions. These subjective insights complement the performance data in illuminating how adaptive tactile feedback principles can enhance attention regulation.

\begin{figure*}
\centering
\includegraphics[width=1\linewidth]{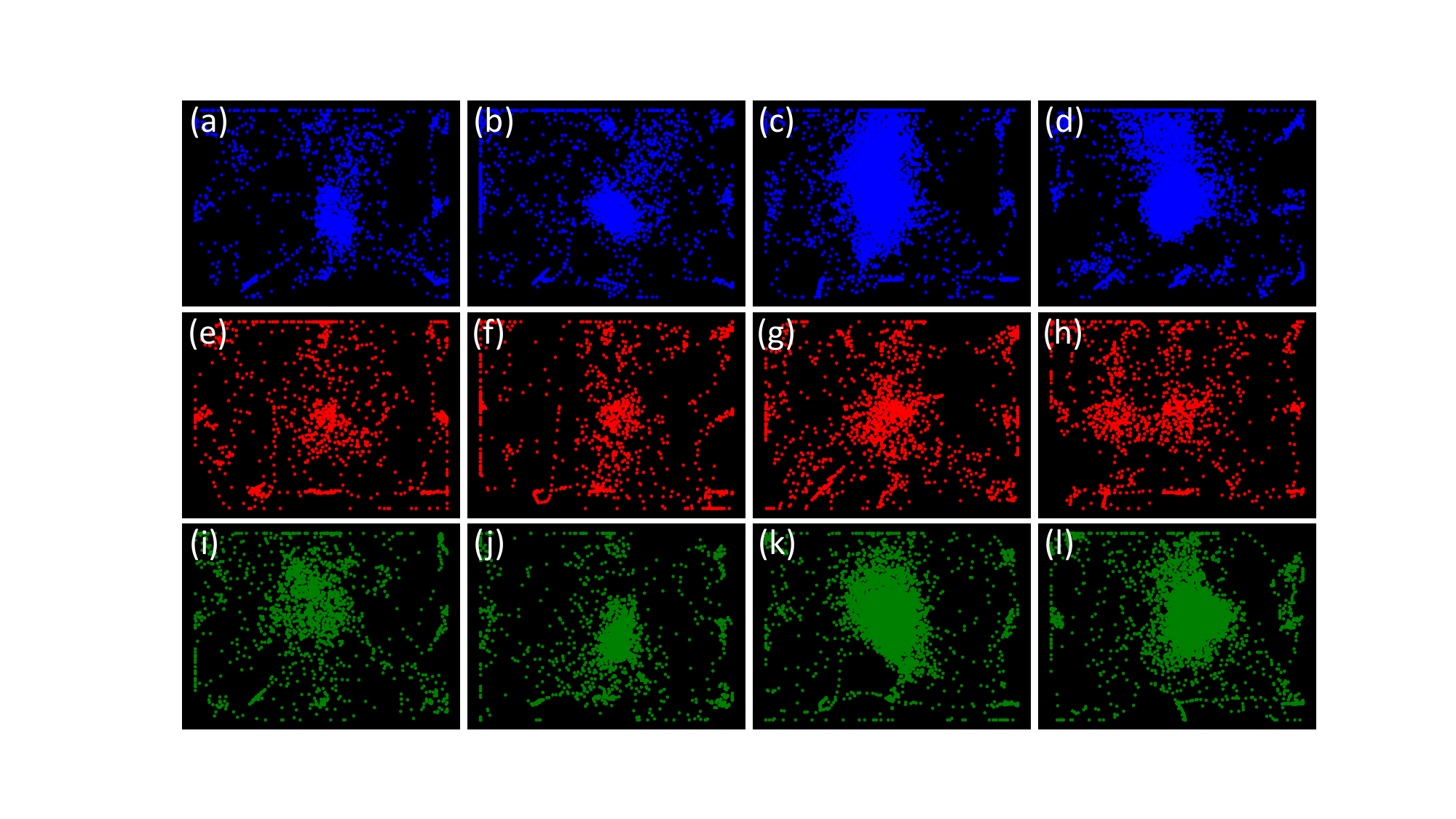}
\caption{Example gaze point cloud of one specific participant under three feedback types in different duration and distraction settings. (a,b,c,d) represent the gaze point cloud with \textit{silence} feedback under short/with distraction, short/no distraction, long/with distraction, and long/no distraction respectively. Similarly, (e,f,g,h) represent the gaze point cloud with \textit{stationary} feedback under such four conditions respectively. (i,j,k,l) represent the gaze point cloud with \textit{filter} feedback under such conditions respectively. }
\label{f3}
\end{figure*}

\subsection*{Gaze Behaviors Indicate Effective Attention Regulation with Eyerofeedback}
\label{sec: R3}
We employed gaze entropy, a commonly used metric in gaze data analysis \cite{shiferaw2019review}, to quantify the participants' attentional focus. Our experiments indicate that gaze entropy was significantly reduced by \textit{filter} eyerofeedback versus \textit{silence} in the long duration conditions (Fig. \ref{f2}(e) and Appendix Fig. S4),  both with distraction (F2,40 = 17.6546, P $<$ 0.0001) and without distraction (F2,40 = 14.2609, P $<$ 0.0001).  
Pairwise comparisons within the long duration setting further revealed significant differences in gaze entropy among the three feedback types in both the no distraction condition (\textit{stationary}: 0.4472 ± 0.9277 (mean ± SD) vs. \textit{silence}: -0.4894 ± 1.0595, P = 0.0063 $<$ 0.01; \textit{filter}: -1.2252 ± 0.7742 vs. \textit{silence}, P = 0.0389 $<$ 0.05; \textit{filter} vs. \textit{stationary}, P $<$ 0.0001) and the distraction condition (\textit{filter}: -0.5402 ± 0.6967 vs. \textit{stationary}: 0.7007 ± 0.5295, P $<$ 0.0001; \textit{stationary} vs. \textit{silence}: -0.3951 ± 0.8576, P $<$ 0.0001). In particular, gaze entropy significantly decreased with \textit{filter} feedback versus \textit{silence}.  Counterintuitively, gaze entropy increased with \textit{stationary} feedback. 
A possible explanation is that continuous tactile cues from \textit{stationary} feedback heighten anxiety and cognitive load since tactile stimuli always happen, which is supported by the higher distraction score felt by participants in \textit{stationary} feedback compared with \textit{filter} feedback in the previous subjective questionnaire.

On the other hand, no significant effects were observed with eyerofeedback (\textit{stationary} or \textit{filter}) in the short duration setting. The differential effects of eyerofeedback observed in long versus short durations were corroborated by the 
interaction between feedback type and duration (F2,40 = 10.1157, P = 0.0003 $<$ 0.001). 
The entropy heatmap in Fig.~\ref{f2}(a) provides an intuitive visualization of these findings. 
By enhancing gaze awareness and regulation, eyerofeedback reduced gaze entropy, aligning with improved response times. \textit{filter} feedback optimally focused gaze, particularly amidst distractions, highlighting the benefits of adaptive tactile feedback principles.

The findings confirm our hypothesis that tactile-bodily gaze mapping via eyerofeedback enhances attention, particularly when demands on sustained focus are high. Eyerofeedback heightens gaze awareness and guides gaze directions, ameliorating attentional lapses. Notably, \textit{filter} feedback outperformed \textit{stationary} feedback in regulating attention. By adaptively triggering tactile cues, \textit{filter} eyerofeedback may mitigate extraneous cognitive load and anxiety associated with continuous \textit{stationary} feedback. Thus, incorporating adaptive principles into tactile biofeedback maximizes benefits on attentional modulation. Refer to Fig.~\ref{f3}, which visually demonstrates concentrated gaze fixations induced by \textit{filter} versus \textit{stationary} or \textit{silence} feedback. In conclusion, rendering gaze patterns on the body optimizes self-monitoring and control of attention through heightened gaze awareness.

\section*{Discussion}
Our investigation examines the potential efficacy of rendering users' gaze patterns on the body to enhance self-awareness of gaze behaviors and regulate attention. 
These findings align with the theory of interoceptive inference \cite{barrett2015interoceptive}, which posits that the human brain can both estimate and regulate critical homeostatic and physiological variables \cite{iodice2019interoceptive}. Within this framework, it is plausible that the brain estimates interoceptive physiological signals that individuals often disregard when not consciously attending to them. By providing real-time feedback of gaze behaviors, the brain's estimation process of these overlooked signals could be strengthened.
Another plausible explanation draws upon neurofeedback theory, which utilizes online presentation of recorded electroencephalography (EEG) in audio/visual formats to assist individuals in consciously controlling their brain waves \cite{marzbani2016neurofeedback}. Analogously, eyerofeedback presents gaze patterns through tactile feedback to help individuals consciously regulate attentive behaviors.

Our observations reveal varying effects of eyerofeedback on human attention under different conditions. In easy (short-duration) tasks, where individuals could consistently maintain focus, eyerofeedback did not significantly impact attention. However, in challenging (long-duration) tasks, where lapses in attention or difficulty sustaining focus over time may occur, eyerofeedback demonstrated an ability to redirect gaze and regulate attention. This was evidenced by faster response times, fewer missed stimuli, and decreased gaze entropy. Furthermore, we found eyerofeedback's influence on attention depended on the applied threshold, particularly with external distractions. Specifically, without distractions, \textit{stationary} eyerofeedback exhibited superior attention regulation over \textit{filter} eyerofeedback. However, with distractions, \textit{filter} eyerofeedback became more effective. We posit distractions may induce anxiety \cite{denkova2010impact}, mitigated through \textit{filter} eyerofeedback's adaptive tactile control. Without distractions, \textit{stationary} eyerofeedback consistently stimulated arousal and attention. But individuals may have habituated to this stimulus during the distraction-filled scenario \cite{mehrabian1977questionnaire}, diminishing its regulatory capacity. In contrast, \textit{filter} eyerofeedback selectively provided reminders, preventing habituation.

Although eyerofeedback significantly influenced human attention only during long-duration tasks, participants reported noticeable subjective effects even in short-duration settings. This indicates eyerofeedback had a substantial impact on users' subjective experiences across all conditions, despite performance improvements being primarily observed in long-duration tasks.
One explanation may be that eyerofeedback effectively delivered regulatory signals to individuals in all scenarios. However, these signals may not have triggered immediate behavioral control in the brain within short timeframes. In other words, the brain potentially received the signals without generating sufficient action to control behaviors in the short-duration tasks \cite{merker2007consciousness}.
Alternatively, eyerofeedback could have impacted attention in short-duration tasks without reaching statistical significance. This may be because performance was already high at baseline \cite{broadhurst1957emotionality} in short-duration tasks. While present, enhancements from eyerofeedback may have been insufficiently large relative to baseline to achieve statistical significance, given the initially elevated performance levels.


Furthermore, we discovered a synchronization effect between response time and gaze entropy. Significant effects on both metrics were observed solely in long-duration tasks, not short ones. This indicates eyerofeedback may enhance attention by regulating and redirecting gaze. Additionally, significant gaze entropy differences occurred only across feedback and duration conditions, unaffected by distractions. This implies that changes in gaze patterns resulted predominantly from eyerofeedback rather than distractions. Such robust evidence bolsters claims that eyerofeedback effectively modulates and guides attention.

One limitation of this study is the relatively small sample size, which restricts generalizability of findings to broader or specific demographic populations. Larger-scale studies would promote greater representation and generalizability across diverse groups. Another limitation involves the current tactile modality for eyerofeedback. Mapping continuous gaze to vibrations on discrete body parts may lose information about gaze smoothness. Intuitively, continuous tactile feedback could better represent gaze motions. However, implementing such a system poses challenges. Full-body wearables may be cumbersome, while smaller wrist devices could impede discerning continuous tactile stimuli \cite{zigler1926tactual}. Even solving these issues, continuous feedback might distract, adversely affecting attention \cite{harman1997distress}. Future work could explore whether continuous tactile mapping elicits different attentional effects compared to discrete stimuli \cite{evans1991tactile}. Additionally, our attention task required fixed screen center focus to measure attention. However, real-world scenarios often involve more complex visual search where eye movements are integral \cite{wolfe2017five}. Moreover, fixed gazing does not guarantee sustained attention, as absent-mindedness may still occur \cite{cheyne2006absent}. Thus, future studies should investigate tactile-bodily-gaze mapping in more intricate scenarios, potentially yielding further insights.

Our findings may inspire new technologies enhancing self-awareness and addressing interoceptive processing issues like ADHD \cite{monastra2006electroencephalographic} and fatigue \cite{windthorst2017heart}. This work significantly contributes to biofeedback literature \cite{biofeedback2008biofeedback}, which has focused on neurofeedback \cite{hammond2011neurofeedback}. As a novel form of biofeedback, eyerofeedback presents intuitive, and comprehensible gaze patterns, compared with neurofeedback's abstract brain visualizations \cite{kimmig2001relationship}. However, despite different implementations, both leverage bio-signals to foster self-awareness. Consequently, eyerofeedback could apply to various neurofeedback scenarios like cognitive training \cite{jiang2017tuning}, rehabilitation \cite{giggins2013biofeedback} and anxiety release \cite{micoulaud2021eeg}. Eyerofeedback's unique benefits and broad applicability make it a promising avenue for integration into interventions and practices improving self-awareness and addressing attention challenges.

\section*{Materials and Methods}

\subsection*{Participants}
We recruited 26 adult participants from a local university (age: 23.5 ± 2.3 years; 6 females). All were right-handed with normal hearing and vision (with eyeglasses). Each participant received 10 USD compensation and was asked to wear headphones during the study. Three participants were excluded for not wearing headphones, and two were excluded due to incomplete data from technical issues, leaving 21 participants for analysis.
We used a within-subject design where each participant completed 12 randomized sessions combining 3 feedback types (\textit{silence}, \textit{stationary}, \textit{filter}) x 2 durations (short, long) x 2 conditions (with/without distraction). Session order randomization ensured variability. The University of California San Diego Institutional Review Board approved the study, and we obtained written informed consent from all participants beforehand.

\subsection*{Attention Task}
We employed a modified three-choice vigilance task (3CVT) to measure attention \cite{meghdadi2021eeg}. Participants were presented with three geometrical shapes: a target upward triangle, non-target downward triangle, and diamond distractor. Each of the ten trials per session displayed one shape followed by a random time interval. Shapes followed a 4:3:3 (target:non-target:distractor) ratio, with order shuffled to prevent bias.
We made two key modifications to the original 3CVT \cite{meghdadi2021eeg}. First, we introduced short (2-5 s) and long (25-35 s) intervals between shapes, manipulating difficulty. Longer durations required extended focus, thus increasing difficulty. Second, rather than random locations, shapes appeared at the center, enabling direct measurement of continuous attention and response times. Gaze deviations from center indicated lapsed attention, thus increasing response times. Participants pressed left/right arrows for target/non-target shapes, allowing the response time measurement.

\subsection*{Tactile Bodily Gaze Map}
The eyerofeedback system delivered vibratory stimuli to users' wrists and ankles based on eye movement directions. We divided the screen into four areas - Upper Left, Upper Right, Lower Left, Lower Right (Fig.~\ref{f1}) - establishing the following mapping: Upper Left $\rightarrow$ Left Wrist, Upper Right $\rightarrow$ Right Wrist, Lower Left $\rightarrow$ Left Ankle, Lower Right $\rightarrow$ Right Ankle. As users shifted gaze across areas, corresponding vibrations were triggered on their body. For instance, gazing upper left induced left wrist vibration, allowing users to sense their eye movements and regulate attention accordingly.

Vibratory tactile stimuli were delivered via 3D printed black wristbands housing vibration motors (Fig. S1 in Appendix). An Arduino Uno \cite{WinNT} controlled the 1 Hz vibration motors from PC commands. Adjustable wristbands using Velcro ensured comfortable wearing on both wrists and ankles.

Eye gaze data was collected using WebGazer \cite{papoutsaki2016webgazer}. Participants underwent pre-study calibration by clicking dots at various screen locations. Notably, we did not capture or save any face images during data collection. The gaze data contained timestamps for synchronization and (x,y) coordinates of eye movements on-screen.

\subsection*{Experiment Apparatus and Instruction}

The experimenter began by introducing the basic procedures of the study and addressing any questions or concerns participants had, ensuring their full comprehension of the information provided in the consent form. Subsequently, the experimenter assisted participants in wearing the four wristbands on their wrists and ankles. To minimize any additional pressure, the experimenter did not observe the participants or the screen during the formal study. However, if participants encountered any issues or had questions during the study, the experimenter was available to provide assistance.

Participants were informed that they would be completing an attention task and would receive tactile stimuli on their wrists and ankles, corresponding to their eye movements. They were instructed that the only way to reduce or eliminate the feedback was to maintain focus on the center of the screen. Prior to the formal study, participants were given an explanation of the attention task and provided with an opportunity to practice, allowing them to become familiar with the task.

After each session, participants were instructed to take a minimum of 1-minute rest and complete a questionnaire (Fig.~\ref{f2}(d)) before proceeding to the next session. If participants felt fatigued from the previous study session, they were encouraged to take a longer rest period. Additionally, participants were required to perform a new eye tracking calibration after each session to minimize potential drift in WebGazer.

It is important to note that participants were instructed to wear the wristbands containing the vibration motors throughout the entire study, even during sessions when the feedback type was set to \textit{silence}. Finally, participants were instructed to prioritize accuracy in the attention task, ensuring they pressed the correct key on the keyboard when a shape appeared, and then attempting to respond as quickly as possible.

\subsection*{Data and Code Availability}

All data and codes needed to reproduce the findings and analysis are available at github: https://github.com/songlinxu/Eyerofeedback.

\bibliography{reference}

\end{document}